\documentstyle[12pt]{article}
\begin{document}
\centerline{\bf Effect of attenuation and dispersion
in the communication channel}
\vskip 2mm
\centerline{\bf on the secrecy of a quantum cryptosystem}
\vskip 3mm
\centerline{S.N.Molotkov}
\vskip 3mm
\centerline{\sl\small
Institute of Solid State Physics, Chernogolovka, 142432 Russia}
\vskip 3mm
\begin{abstract}
The effects of dispersion in the communication channel on the
secrecy of a quantum cryptosystem based on single photon states with 
different frequencies are studied. A maximum communication channel length
which can still ensure the secrecy of the key generation procedure is found.
\end{abstract} 
\vskip 3mm

The main purpose of cryptography is to allow the exchange of secret
information among two or more legitimate users. The basic element
of every cryptosystem is the key, i.e. a random sequence of units and
zeros used to code the messages [1]. The communication can be shown to be
absolutely secret if the key length equals the message length and the key
is used only once [2]. Therefore, the major task is to ensure the secrecy
of the key distribution procedure among the legitimate users.
In the standard cryptosystem there is no fundamental principle
which could guarantee the detection of any eavesdropping attempt
at the key distribution stage; thus the cryptosystem secrecy is based
on the key complexity rather than fundamental laws of nature [1].

On the other hand, quantum cryptography provides a key distribution 
procedures where the possibility of detecting any eavesdropping attempt
is guaranteed by the fundamental laws of quantum mechanics.

As a rule, the secrecy of quantum cryptosystems is proved for
ideal communication channels. However, the imperfections of a
communication channel should generally reduce the secrecy of the
key generation procedure so that any practical cryptosystem should
carefully take into account the actual properties of the communication 
channel employed.

Recently, several new quantum cryptosystems have been proposed [3--9].
One of these systems, based on phase coding and employing a 30 km long 
fiber line as an interferometer arm has been realized experimentally [9].

In the paper [10] a quantum cryptosystem based on the EPR 
(Einstein--Podolsky--Rosen)
effect for biphoton states has been proposed. Actually, a similar scheme
can be implemented with single-photon states, which is much simpler from 
the experimental point of view since it does not involve generation of 
biphoton states (e.g., with the parametric down-conversion).
{\sl In addition, such a scheme should be much more stable, since it does
not employ a long-arm interferometer.}

The secrecy of this cryptosystem (detection of eavesdropping attempts) 
is based on the fundamental time--energy uncertainty relation. The scheme
proposed in [10] did not take into account attenuation and dispersion in the 
communication channel which could severely hamper its practical realization.
The purpose of the present paper is to find out the conditions under which
it is possible to ensure the secrecy of the cryptosystem in the
presence of attenuation and dispersion.

Let us first describe the key generation procedure which does not
involve the biphoton states. The user A sends at random into the 
communication channel (optical fiber) to user B on of the following 
three single-photon states: one of the two states with narrow frequency
spectra centered around well defined frequencies $\omega_1$ and $\omega_2$
(frequency spectra widths $\sigma_1, \sigma_2 \ll |\omega_1 - \omega_2|$)
or a well localized in time (correspondingly, with a wide frequency 
spectrum of width $\sigma_{\infty}$) at a carry frequency 
$\omega_0\approx\omega_{1,2}$. Actually, the optical fiber transparency
window corresponds to the wavelength $\lambda\approx 1.3$ $\mu$m 
(frequencies $\omega_{0,1,2}\approx 10^{15}$ s$^{-1}$).

According to the fundamental time--energy uncertainty relation, 
sending a signal with a well-defined frequency $\omega_1$ or 
$\omega_2$ means that the moments of times when the photon
leaves user A ($t_A$) and is registered by user B ($t_B$) 
exhibit a large scatter $\Delta t_{A,B} \ge 1/\sigma_{1,2}$.

If a broad-spectrum photon is emitted, although 
the associated uncertainty in frequency 
$\sigma_{\infty}$ is large,
the corresponding state can be prepared rather fast and the
photon emission and detection times can be measured with high 
accuracy ($\Delta t\approx 1/\sigma_{\infty}\rightarrow0$  
if $\sigma_{\infty}\rightarrow\infty$).

To register a photon, user B choses randomly and independently of user A
in each measurement either one of the narrow-band photodetectors with
central frequencies $\omega_1$ and $\omega_2$ and bandwidths 
$\gamma_{1,2}\approx\sigma_{1,2}$, or a wide-band photodetector with 
the central frequency $\omega_0$ and bandwidth
$\gamma_{\infty}\approx\sigma_{\infty}$. The frequency separation
$\delta\omega_{12}=|\omega_1-\omega_2|$ should not be less 
than the sum $\sigma_1 + \sigma_2$ if the photons with
frequencies $\omega_1$ and $\omega_2$ should be distinguished. 
For a gaussian spectrum the inequality
$\delta\omega_{12} > 3(\sigma_1+\sigma_2)$ is sufficient.

Measurements with a narrow-band photodetector allow to distinguish
between $\omega_1$ and $\omega_2$, but they
cannot be performed in a time shorter than $1/\delta\omega_{12}$
(the same is also true for the minimal time required {\it to
prepare these states}).

Measurements with a wide-band photodetector can be completed
during the time interval 
$\Delta t_{\infty}\approx1/\sigma_{\infty}\rightarrow 0$, 
but they cannot determine the photon energy with the accuracy
better than $\sigma_{\infty}$. Users A and B choose the cryptosystem 
parameters to satisfy the inequality
\begin{equation}
\Delta t_{\infty}\ll\Delta t_{12}.
\end{equation}

After a series of measurements user B announces through a public channel
which type of the photodetector (wide- or narrow-band) was used in each
measurement, but does not disclose which particular frequency 
$\omega_1$ or $\omega_2$) was used in the case of a narrow-band
photodetector. Those measurements where the photodetector did not fire
or the photodetector type differed from the type of the single-photon
signal, are discarded. The remaining measurements where narrow-band 
photodetectors were used yield a random sequence of zeros and units
($\omega_1$ corresponds to zero and $\omega_2$ to unit)
shared by the two users which can be used as a key.
The probability of an error (e.g., obtaining zero instead of unit)
is negligibly small if $\delta\omega_{12}>3(\sigma_1+\sigma_2)$. 
To correct the key one can use a privacy amplification scheme proposed by
Bennett et al [11].

Measurements where short pulses were used (i.e., the photon
emission and registration times are known with high accuracy)
allow to detect any eavesdropping attempt. For all such measurements 
users A and B announce through a public channel the photon
emission ($t_A$) and registration ($t_B$) times. Then the expected 
delay time $t_A-t_B=const$ (to within $\Delta 
t_{\infty}\approx1/\sigma_{\infty}\rightarrow 0$) is calculated 
from the known line length. Any systematic deviation of 
$t_A-t_b$ from the expected delay time means the presence 
of an eavesdropper. Indeed, to extract the information about the key,
the eavesdropper should be able to distinguish between
$\omega_1$ and $\omega_2$ (0 or 1); therefore, he should employ
narrow-band photodetectors. Such measurements (as well as {\it
preparation} of narrow-band signals centered around 
$\omega_1$ and $\omega_2$ to be sent by the eavesdropper to user B) 
cannot be performed faster than in 
$\Delta t_{12}\approx1/\delta\omega_{12}\gg\Delta t_{\infty}$. 
The eavesdropper will unavoidably run into the situation
where user A sent a short signal, while the eavesdropper uses a
narrow-band photodetector (since the user A chooses the type of signal
he sends to user B at random) and re-sends to user B a signal with a 
well-defined frequency. The eaves dropper must re-send the photon to
user B since otherwise this measurement will be discarded because
the photodetector would not fire. Therefore, a systematic deviation
of $t_A-t_B$ from the expected delay time 
by not less than $\Delta t_{12}\approx 1/\delta\omega_{12}$, which is
much larger than the accuracy with which the delay time $t_A-t_B$ is 
known. 

Up to this moment we did not take into account attenuation and dispersion
of the quantum communication channel. Practically, a fiber cable is used 
as channel which implies that short pulses sent by user A would experience
broadening (signal width at the receiving end of the line is expected
to be enhanced) so that the scatter in the photon registration time
by user B is increased simplifying the task of the eavesdropper and
reducing the cryptosystem security.

Our purpose is find out the relationships between the single-photon states
parameters  $\sigma_{1,2}$, $\delta\omega_{1,2}$, $\sigma_{\infty}$, 
and the communication channel attenuation and dispersion which still allow
a secret key distribution procedure.

Let the user A prepare a single-photon state at the input of the
communication channel (point $x=0$) with the spectral width 
$\sigma$ (which is one of $\sigma_{1,2,\infty}$) and the carry frequency
$\omega_0$ (for definiteness we assume that 
$\omega_0=\omega_{1,2}$, although this is not essential)
\begin{equation}
E(0,t)=\frac{\textstyle 1}{\textstyle (\pi\sigma^2)^{1/4}}
\int_{0}^{\infty}
\exp{\left\{- \frac{\textstyle (\omega-\omega_0)^2}{\textstyle 
2\sigma^2}\right\}} \exp{(-i\omega t)}d\omega
\end{equation}

The effective pulse duration at the channel input is
\begin{equation}
(\Delta t_A)^2=
\int_{0}^{\infty}(t-\overline{t})^2
|E(0,t)|^2dt=\frac{\textstyle 1}{\textstyle 2\sigma^2},
\end{equation}
\begin{displaymath}
\overline{t}=
\int_{0}^{\infty}t|E(0,t)|^2dt,
\end{displaymath}
and its spectral width is
\begin{equation}
(\Delta\omega_A)^2=
\int_{0}^{\infty}(\omega-\overline{\omega})^2
|E(0,\omega)|^2d\omega=\sigma^2/2,
\end{equation}
\begin{displaymath}
\overline{\omega}=
\int_{0}^{\infty}\omega|E(0,\omega)|^2d\omega,
\end{displaymath}
\begin{displaymath}
E(0,\omega)=\int_{0}^{\infty} E(0,t)\exp{(i\omega t)}dt
\end{displaymath}

Actually, even for a short pulse with  
$\overline{t}\approx 10^{-12}$ s the spectral width is relatively
small (carry frequency $\omega_0\approx 10^{15}$ s$^{-1}$), 
so that only quadratic terms can be retained in the expansion of the
wavevector as a function of frequency [12,13]:
\begin{equation}
k(\omega)=k_0+\alpha(\omega-\omega_0)+\beta(\omega-\omega_0)^2,
\end{equation}
where $\alpha$ and $\beta$ are generally complex constants, 
their real and imaginary parts describing dispersion and 
attenuation, respectively. 
Let us first assume that the attenuation is absent.
The signal (2) at the point $x$ (user B) takes the form
\begin{equation}
E(x,t)=
\frac{\textstyle 1}{\textstyle 2\sqrt{\pi}}
\frac{\textstyle 1}{\textstyle \sqrt{\sigma_{0}^{2}-i\beta x}}
\exp{\left\{ 
-\frac{\textstyle (\alpha x-t)^2}
{\textstyle 4(\sigma_{0}^{2}-i\beta x)} 
\right\}},\mbox{ }\sigma^{2}_{0}=
\frac{\textstyle 1}{\textstyle 2\sigma^2}
\end{equation}
The field intensity observed by user B is
\begin{equation}
|E(x,t)|^2=
\frac{\textstyle 1}{\textstyle 4\pi(\sigma_{0}^{2}+\beta^2 x^2)}
\exp{\left\{ 
-\frac{\textstyle (\alpha x-t)^2}
{\textstyle 2\sigma^{2}(x)} \right\}},\mbox{ }
\sigma^2(x)=\frac{\textstyle \sigma_{0}^{4}+\beta^2x^2}
{\textstyle 2\sigma^{2}_{0}}
\end{equation}
The effective spectral width of the signal at point $x$ remains the same as 
at the channel input (remember that the attenuation is not taken into account)
\begin{equation}
(\Delta\omega_A)^2=(\Delta\omega_B)^2=
\int_{0}^{\infty}
(\omega-\overline{\omega})^2|E(0,\omega)|^2d\omega=
\int_{0}^{\infty}
(\omega-\overline{\omega})^2|E(x,\omega)|^2d\omega
\end{equation}
The effective pulse duration at the receiving end of the channel
is increased by a factor of $(1+\beta^2 x^2\sigma^4)$ 
\begin{equation}
(\Delta t_B)^2=
\int_{0}^{\infty}
(t-\overline{t})^2|E(x,t)|^2dt=
\frac{\textstyle 1}{\textstyle 2\sigma^{2}(x)}=
\frac{\textstyle 1}{\textstyle 2\sigma^2}(1+\beta^2 x^2\sigma^4),
\end{equation}
This time $\Delta t_B$ is the mean time taken by the wave packet
to pass through the point $x$, while the inequality
\begin{equation}
\Delta\omega_B\Delta t_B\ge 
\sqrt{1+\beta^2 x^2\sigma^4}
\end{equation}
is a variety of the Mandelstam--Tamm inequality [14].
Equation (10) describes the statistics of (potential) measurements
performed on a particle, rather than an actual measurement of a photon
observable so that it cannot be interpreted as a time--energy uncertainty
relation relevant to a real act of measurement (which is described by
the Bohr uncertainty principle; see a detailed discussion in the paper
by Krylov and Fock [16]). We shall adhere to the orthodox point of view
assuming the time--energy uncertainty relation is a fundamental law of nature
(various point of view are discussed in a review article by
Dodonov and Man'ko [17]). The average time taken by the wave packet to pass 
through the point $x$ has nothing to do with the measurement time 
$\delta t$ which is arbitrarily chosen by the experimentalist.
The Bohr uncertainty relation is applicable to a real act of measurement
(e.g. passage of a particle through a device shutter which unavoidably changes
the particle energy in an uncontrollable way) 
\begin{equation}
\Delta E\Delta t\ge 1,
\end{equation}
where $\Delta E$ is the scatter of measured energy [17]. 
Unlike the Mandelstam-Tamm relation which is derived from the evolution
governed by the Schr\"{o}dinger equation [14], 
the Bohr uncertainty relation is actually a postulate since 
the measurement act cannot be described by the Schr\"{o}dinger equation.

Thus, to register a photon with the spectrum width $\sigma_{\infty}$ 
user B should open the shutter for a time interval at least
$(1+\beta^2x^2\sigma_{\infty}^{4})^{1/2}/\sigma_{\infty}$ long.
Of course, user B could open the shutter for arbitrarily short time,
but in that case he would not be able to systematically detect a photon:
Although in some rare measurements he would still register a photon
even if the shutter were only open during $\delta t\rightarrow 0$,
the fraction of such measurements should tend to zero or otherwise
the Bohr time--energy uncertainty relation  [15] would be violated.

Let us now find out when the quantum cryptosystem still remains secret,
i.e. the scatter in the short pulse registration times by user B $\Delta t_B$ 
should be substantially less than the time taken by the eavesdropper
(at the location $x$ somewhere between A and B) to register the photon 
with a narrow-band photodetector, which, as described above, could not be 
shorter than
\begin{equation}
\Delta t_E\ge 
\frac{\textstyle (1+\beta^2x_{E}^{2}\sigma_{1,2}^{4})^{1/2}}
{\textstyle \delta\omega_{12}}
\end{equation}
The inequality 
\begin{equation}
\Delta t_{B}\ll\Delta t_E
\end{equation}
imposes the following limit on the channel length:
\begin{equation}
\frac{\textstyle (1+\beta^2x^2\sigma_{\infty}^{4})^{1/2}}
{\textstyle \sigma_{\infty}}\ll
\frac{\textstyle (1+\beta^2x_{E}^{2}\sigma_{1,2}^{4})^{1/2}}
{\textstyle \delta\omega_{12}}
\end{equation}
The worst situation with respect to the system secrecy occurs
if the eavesdropper is located close to the user A ($x_E\approx 0$).
In that case the eavesdropper is not affected by the pulse broadening.
Therefore, the maximum channel length is  
\begin{equation}
x_B\le\frac{\textstyle 1}
{\textstyle \delta\omega_{12}\sigma_{\infty}\beta}
\end{equation}
Thus, the smaller is the frequency separation between the information-carrying
signals $\omega_1$ and $\omega_2$, the shorter is the reference
pulse (the wider is its frequency spectrum), and the lower is the
dispersion quadratic coefficient, the larger is the allowed quantum 
communication channel length still preserving the system secrecy.
However, this inequality does not impose any absolute restrictions
on the channel length. Formally, the channel length can be made
arbitrarily large at the price of reducing the frequency separation
$\delta \omega_{12}=|\omega_1-\omega_2|$.

Let us now make some numerical estimates. For a frequency separation
$\delta\omega_{12}=|\omega_1-\omega_2|\approx 10^9$ Hz corresponding
to the linewidth of a rather average semiconductor laser, 
the short pulse duration of 1 ps
($\sigma_{\infty}\approx 10^{12}$ Hz), and a typical quadratic dispersion
coefficient $\beta\approx 1$ ps$^2$/km [18], one has for the allowed 
channel length
\begin{displaymath}
x_B\le 
\frac{\textstyle 1}{\textstyle 10^9\cdot10^{12}\cdot (10^{-12})^2}\mbox{ }
[\mbox{km}]
\approx 10^3\mbox{ } \mbox{km},
\end{displaymath}

The role of attenuation reduces to the following two effects.
First, the fraction of measurements where the photodetector
employed by user B did not fire is increased. This effect
reduces the system efficiency but does not affect its secrecy.
The second effect is the renormalization of dispersion.
Now the restriction on the channel length becomes
\begin{equation}
\frac{\textstyle 
(1+\sigma^{2}_{\infty}\beta_{im}x)^2+\beta_{re}^2x^2\sigma^{4}_{\infty}}
{\textstyle (1+\sigma^{2}_{\infty}\beta_{im}x)^2} 
\le
\frac{\textstyle \sigma^{2}_{\infty}}{\textstyle \delta\omega_{12}};
\end{equation}
for weak attenuation
\begin{equation}
x_B\le\frac{\textstyle 1}
{\textstyle \sqrt{\beta_{im}^{2}+\beta_{re}^{2}}\delta\omega_{12}
\sigma_{\infty}},
\end{equation}
where $\beta_{re}$ and $\beta_{im}$ are the real and imaginary parts 
of the dispersion quadratic coefficient.

The author is grateful to S.V.Iordansky,
S.S.Nazin, and S.T.Pavlov for fruitful discussions during this study. 
This work was supported by the Russian Fond for Fundamental 
Research (grant No 96-02-19396).


\begin{thebibliography}{99}
\bibitem{1}M.E.Hellman, Sci. Amer., {\bf 241}, 130 (1979);
G.J.Simmons, The Math. Intelligencer, {\bf 1}, 233 (1979).
\bibitem{2}C.E.Shannon, Bell Syst. Techn. J. 28, 657 (1949).
\bibitem{3}A.K.Ekert, Phys. Rev. Lett. {\bf 67}, 661 (1991).
\bibitem{4}C.Bennett, Phys. Rev. Lett. {\bf 68}, 3132 (1992).
\bibitem{5}C.H.Bennett, G.Brassard, N.D.Mermin, 
Phys. Rev. Lett. {\bf 68}, 557 (1992).
\bibitem{6}A.K.Ekert, J.G.Rarity, P.R.Tapster, G.M.Palma, 
Phys. Rev. Lett. {\bf 69}, 1293 (1992).
\bibitem{7}R.J.Hughes, D.M.Alde, P.Dyer, G.G.Luther, G.L.Morgan, M.Schauer,
Contemporary Phys. {\bf 36}, 149 (1995).
\bibitem{8}S.J.D.Phoenix, P.D.Townsend, Contemporary Phys. {\bf 36}, 165 
(1995).
\bibitem{9}C.Marand, P.D.Townsend, Optics Lett. {\bf 20}, 1695 (1995).
\bibitem{10}S.N.Molotkov, S.S.Nazin, Pis'ma ZhETF, {\bf 63}, 882 (1996).
\bibitem{11}C.H.Bennett, F.Bessette, G.Brassard, L.Salvail, J.Smolin,
J. Cryptology, {\bf 5}, 3 (1992).
\bibitem{12}J.D.Franson, Phys. Rev., {\bf A45}, 3126 (1992).
\bibitem{13}J.Jeffers, S.Barnett, Phys. Rev., {\bf A47}, 3291 (1993).
\bibitem{14}L.I.Mandelstam, I.E.Tamm, Isvestya USSR Acad. Sci, ser. phys.,
{\bf 9}, N 1/2, 122 (1945).
\bibitem{15}N.Bohr, Selected Scientific Papers, vol 2, 675pp, 
Moscow, Nauka, 1971.
\bibitem{16}N.S.Krylov, V.A.Fock, ZhETF, {\bf 17}, 93 (1947);
V.A.Fock, ZhETF, {\bf 42}, 1135 (1962).
\bibitem{17}V.V.Dodonov, V.I.Man'ko, Trudy FIAN, {\bf 183}, 52 (1987).
\bibitem{18}G.P.Argawal, {\it Nonlinear Fiber Optics}, ch. 3, (1989),
Academic Press, Inc., Harcourt Brace Jovanovich Publishers.
\end{thebibliography}
\end{document}